\documentclass[12pt,preprint]{aastex}

 1
\newcommand{\ie}{{\it i.e.}}

\newcommand{\Msun}{M_\odot}

\newcommand{\lta}{\lower.5ex\hbox{\ltsima}}
\newcommand{\gta}{\lower.5ex\hbox{\gtsima}}
\newcommand{\ltsima}{$\; \buildrel < \over \sim \;$}
\newcommand{\lsim}{\lower.5ex\hbox{\ltsima}}
\newcommand{\gtsima}{$\; \buildrel > \over \sim \;$}
\newcommand{\gsim}{\lower.5ex\hbox{\gtsima}}

\newcommand{\upm}{{\stackrel{\raise.5ex\hbox{$m$}}{\lower.2ex\hbox{.}}}}
\newcommand{\upd}{{\stackrel{\raise.5ex\hbox{$d$}}{\lower.2ex\hbox{.}}}}

\newcommand{\gs}{\mathrel{\raise1.16pt\hbox{$>$}\kern-7.0pt
\lower3.06pt\hbox{{$\scriptstyle \sim$}}}}
\newcommand{\ls}{\mathrel{\raise1.16pt\hbox{$<$}\kern-7.0pt
\lower3.06pt\hbox{{$\scriptstyle \sim$}}}}
\newcommand{\prosima}{$\; \buildrel \propto \over \sim \;$}   
\newcommand{\simgt}{\lower.5ex\hbox{\gtsima}}
\newcommand{\simlt}{\lower.5ex\hbox{\ltsima}}
\newcommand{\simpr}{\lower.5ex\hbox{\prosima}}

\newcommand{\eeq}{\end{equation}}
\newcommand{\eeqa}{\end{eqnarray}}



\begin{document}

\title{Star Formation in Space and Time: \\
    Taurus-Auriga}

\author{Francesco Palla\altaffilmark{1} and Steven W. Stahler\altaffilmark{2}}

\altaffiltext{1}{Osservatorio Astrofisico di Arcetri, Florence}
\altaffiltext{2}{Berkeley Astronomy Department}

\slugcomment{submitted to {\it The Astrophysical Journal}, August 15, 2002}
%\received{---------------}
%\accepted{---------------}

\begin{abstract}

To understand the formation of stellar groups, one must first document 
carefully the birth pattern within real clusters and associations. In this 
study of Taurus-Auriga, we combine pre-main-sequence ages from our own 
evolutionary tracks with stellar positions from observational surveys. Aided 
by the extensive, millimeter data on the molecular clouds, we develop a 
picture of the region's history. Star formation began, at a relatively low 
level and in a spatially diffuse manner, at least $1\times 10^7$~yr in the 
past. Within the last few million years, new stars have been produced at an 
accelerating rate, almost exclusively within a confined group of striated 
cloud filaments.

The gas both inside and around the filaments appears to be in force balance.
Thus, the appearance of the filaments is due to global, quasi-static
contraction of the parent cloud material. Gravity drives this contraction and
shock dissipation mediates it, but the internal motion of the gas does not
appear to be turbulent. The accelerating nature of recent star formation means
that the condensation of cloud cores is a threshold phenomenon, requiring a
minimum background density. Other, nearby cloud regions, including Lupus and
Chamaeleon, contain some locales that have attained this density, and others
that have not. In the latter, we find extensive and sometimes massive molecular
gas that is still devoid of young stars. 

\end{abstract}


\section{Introduction}

One basic characteristic of stars is that they form not as isolated objects, 
but in populous groups within molecular clouds and cloud complexes. This  
observational fact raises other basic issues of a theoretical nature. What     
supports a massive cloud against its self-gravity {\it before} the production 
of an interior stellar group? How do individual dense cores, each capable of   
forming single and binary stars, arise within this kinetic, and perhaps
turbulent, medium? Which properties of the large parent cloud determine
whether it will spawn a bound cluster, T association, or expanding OB
association? 
 
Answering all these questions will require a broader and deeper understanding  
than is currently at hand, one that connects the birth of individual stars to 
the growth and evolution of clouds on a multi-parsec scale. Any such theory,
if it is to be quantitative, must be based on detailed information concerning
the earliest stages of existing groups. That is, we should first ascertain
the actual pattern of stellar births in observed clusters and associations.
To address this issue, we began by constructing a set of pre-main-sequence  
evolutionary tracks (Palla \& Stahler 1999; hereafter Paper I). These covered
masses from 0.1 to 8 $\Msun$, \ie, from close to the brown-dwarf regime to
the upper mass limit of the pre-main-sequence phase itself (Palla \& Stahler   
1990). The tracks can be used to assign contraction ages to any young star    
that has reliable values of effective temperature and luminosity. A compilation
of ages then represents the star formation history of the region of 
interest.

Paper I applied this method to the Orion Nebula Cluster. Utilizing the
extensive database of Hillenbrand (1997), we found that star formation began
at a relatively low rate at least $10^7$~yr ago, then increased markedly within
the last $2\times 10^6$~yr. In a subsequent contribution (Palla \& Stahler    
2000; hereafter Paper II), we examined seven additional, nearby groups. Most
exhibited a similar pattern of activity as Orion: a slow rise followed by
rapid acceleration through the current epoch. 

This result has recently been questioned by Hartmann (2001), who claims that
observational errors in effective temperatures and luminosities are so large
that they preclude the inference of any detailed history at all. Hartmann
further contends that the data from T associations are consistent with a rapid 
burst of star formation in the recent past, and that apparently older stars   
have simply been assigned erroneously low luminosities. This view is
contradicted, however, by the HR diagrams in Papers I and II. If the
observational errors were indeed large enough to ``age" a substantial portion
of stars, they would also give spuriously {\it high} luminosities to an equal 
number of other objects. Many of these would appear above the birthline in the
HR diagram, which is not the case. In his statistical analysis, Hartmann avoids
this difficulty by ignoring the birthline and allowing arbitrarily high
luminosities for his sample population (see his equation (2) and Figure 2).   
 
Assuming that our findings are robust, many systems are increasing their
production of stars to the present time. This intriguing result naturally
prompts additional questions. Where are the clouds that are not yet actively
forming stars? Within a given cloud, what ends the acceleration?

A partial answer to the second point is available from the earlier studies. In 
the volume surveyed by Hillenbrand, \ie, within several pc of the Trapezium and
in front of the massive, Orion~A cloud, molecular-line studies show that there
are no dense cores (Bergin 1996). Hence, new stars are no longer forming in 
this region, and their production rate must overturn sharply. We could not 
resolve this rapid decline with our technique. However, one system from Paper 
II did show a decline, albeit over a longer time scale. This was 
Upper~Scorpius, an OB association with very low gas content (de Geus, 
Bronfman, \& Thaddeus 1990).  In an interesting, related study, Dolan \& 
Mathieu (2001) have documented the star formation history of the 
$\lambda$~Orionis region. They found the star formation rate to be 
accelerating in unperturbed cloud material well removed from the massive 
stars, but to have fallen to a low level closer to these objects, where the 
gas has largely been dispersed.

The manner in which accelerating star formation ends within a low-mass
T~association is less clear. The results concerning $\lambda$~Orionis
illustrate, in any case, the value of probing both the temporal and {\it
spatial} pattern of activity. In the present paper, we apply this philosophy   
to the Taurus-Auriga association. This region has, of course, been thoroughly 
studied over many years. Stars have been catalogued through a number of deep
surveys covering the optical, infrared, and X-ray regimes (Kenyon \& Hartmann
1995; Brice\~no et al. 1998; Luhman 2000; K\"onig, Neuhauser, \& Stelzer 2001).
The stellar population is located within a cloud complex that has itself
been mapped systematically in a number of CO isotopes (Kawamura et al. 1998;
Onishi et al. 1996; Dame, Hartmann, \& Thaddeus 2001), and more sparsely in 
other, high-density tracers (Benson \& Myers 1989; Pound \& Bally 1991; Saito
et al. 2001). Thus, we inherit a rich trove of data for assessing the star
formation pattern. 
 
We describe this pattern in \S 2, below, and show how activity has become
centrally concentrated with time. We then focus on those stars {\it   
outside} the central region, to confirm that they are bona fide members of
the association. Section 3 discusses the implications of our findings   
with regard to the basic questions posed earlier.  We also compare our results 
with recently proposed dynamical models of stellar group formation. Finally,  
in \S 4, we indicate fruitful directions for research in the near future.  

\section{The Star Formation Pattern}

\subsection{Previous Studies}

The most striking aspect of the Taurus-Auriga association is the striated
morphology of the parent cloud gas. This basic feature is apparent even in
optical photographs of the region, as was stressed early last century by 
Barnard (1927). Molecular-line studies, including those cited above, have
since revealed the full extent of the gaseous structures. With its linear 
extent of 30~pc and total mass estimated at $2\times 10^4\ \Msun$ (Ungerechts
\& Thaddeus 1987), the complex nearly attains the status of a giant molecular  
cloud. How did this gas accumulate, and how is it producing stars today?

G\'omez de Castro \& Pudritz (1992) boldly addressed both questions in an
interesting theoretical investigation. Using the proper motion study of Jones
\& Herbig (1979), they pointed out that the young stars appear to be
streaming in a direction that is parallel to that of the filaments.
G\'omez~de~Castro \& Pudritz also stressed the wavelike appearance of the
ambient magnetic field, which runs roughly perpendicular to the orientation
of the filaments (Scalo 1990). They explained all these findings by
postulating that the cloud complex formed through the magnetic Parker
instability (Parker 1966).  Gas flowing down buckled field lines crashes into
material already in the plane, creating local concentrations of streaming gas
(the dark clouds), while radiating much of the collision energy in the form
of Alfv\'en waves. They attributed the filaments themselves to density    
enhancements produced by such waves when they attain mildly nonlinear
amplitude. Finally, G\'omez~de~Castro \& Pudritz compared the H$\alpha$
luminosity functions in two stellar groups located at opposite ends of the
complex. As these functions were similar, they surmised that the complex as a
whole is forming stars coevally.

More recently, a very different account was offered by Ballesteros-Paredes
et al. (1999). These authors noted that the molecular gas in Taurus-Auriga is
spatially correlated with HI, as seen through 21-cm surveys. Other researchers
(e.g., Andersson 1993) have interpreted this result as indicating the presence
of an atomic envelope surrounding the molecular complex. Ballesteros-Paredes
et al. emphasized that the HI line profiles are both significantly broader than
those of $^{12}$CO, exhibit a pronounced asymmetry, and are shifted in their
peak velocity. They therefore speculated that the present-day molecular
filaments were created through the collision of high-speed flows in the atomic
gas. The resulting dense, postshock configuration could undergo prompt  
gravitational collapse and initiate a burst of star formation throughout the
region (see also Hartmann et al. 2001). 

A point not emphasized by previous authors is the spatial distribution of the
stars themselves. Two facts ensure that such a study is relevant for the
question of stellar birth.  First, the extensive survey work by numerous
researchers has now provided a nearly complete population census within the
association, at least to a limiting $V$-magnitude of 17 (Brice\~no et al 1999).
Note that the observational criteria here are not limited as before to
emission-line diagnostics, so that the investigations fully account for
weak-lined and post-T~Tauri objects.

Apart from the completeness issue, the current positions of the stars would  
be of limited interest if the objects could drift far from their birth 
sites.  However, the stars must have been born with the same velocity as the
local gas. Observations indicate that this equality still holds. Specifically,
the difference between stellar and cloud radial velocities is at most about 
2~km~s$^{-1}$ (Hartmann et al 1986; see also below). Even objects traveling
on ballistic trajectories with this speed would cover only a few pc in several
Myr, which is the lifetime of most association members. The stellar 
distribution therefore does reflect conditions at birth, a point to which we 
return presently. 

G\'omez et al (1993) pointed out that Taurus-Auriga stars are more clumped
spatially than a random distribution throughout the parent clouds. Indeed,
the median nearest-neighbor separation is only 0.3~pc, not much greater than
the diameter of a star-forming dense core (Myers \& Benson 1983). G\'omez et al
identified six distinct groups, each containing about 15 individual stars or
binaries, and ranging in size from 0.5 to 1.0~pc. The remaining stars, some   
30~percent of the total, have a more uniform distribution. Interestingly,
G\'omez de Castro \& Pudritz (1992) had previously speculated that star 
formation in Taurus-Auriga proceeded through individual clusters. They based
this idea on the observed distribution of molecular clumps, which is far more
weighted to higher masses than the visible stars. Thus, if all clumps produce
stars with comparable efficiency, then the most massive should yield multiple
objects.   

\subsection{New Findings}

We base our own investigation on a sample of 153 stars, drawn from a number
of sources in the literature. Most are from the compilation of Kenyon \&
Hartmann (1995), who in turn utilized and supplemented earlier studies in the 
optical, near-infrared, and X-ray regimes. The membership list of Kenyon \&
Hartmann is complete to \hbox{$V\,\sim\,15\,{\rm mag}$}. Within the L1495E
cloud, both Strom \& Strom (1994) and Luhman \& Rieke (1998) conducted deeper
surveys which doubled the number of known members in this area. Brice\~no et 
al (1998) focused on very low-mass objects in both L1495E and other regions, 
uncovering 9 new stars. Both the 2MASS project (Luhman 2000) and the
large-scale, optical CIDA survey (Brice\~no et al. 1999) yielded additional
objects. Finally, we have added 13 stars from the ROSAT study of Wichmann et
al. (2000). 
 
All told, these various sources gave us 212 member stars. Many, however, do not
have published luminosities and effective temperatures, generally because they
are more deeply embedded Class~0 or I sources. Those that do, a total of 153
objects, can be placed in the HR diagram. For this subset, we were able to 
assign contraction ages based on our own pre-main-sequence tracks (Paper I).
The zero point of these ages corresponds to the stellar birthline.
\footnote{We have compiled essential data for all 212 stars in a table, 
available at
http://www.arcetri.astro.it/$\sim$palla/taurus/table. This table includes
pre-main-sequence ages for the 153 members with published luminosities and
effective temperatures.} 

Figure 1 summarizes our results for the spatial {\it and} temporal
distribution of all 153 stars. Here, we have binned the objects into three age
groups, as indicated. We have also displayed within each panel the
$^{12}$CO~{$(J =1\rightarrow 0)$} contours of Dame et al (2001). The lowest  
contour is for \hbox{$\int\!T_A\,dV\,=\,2\ {\rm K}\ {\rm km}\ {\rm s}^{-1}$},
corresponding to a gas column density of  
\hbox{$N_H \,=\,4\times 10^{20}\,{\rm cm}^{-2}$}, or 
\hbox{$A_V \,=\,0.3\ {\rm mag}$}. Scanning the panels from  
left to right gives a global view of the progress of star formation throughout
the entire complex. 

The five oldest stars in our sample all have ages of $2\times 10^7$~yr. From  
this time in the past until 4~Myr ago, there were numerous stellar births, but
they were widely distributed in space. On the other hand, a few discrete   
centers of activity are apparent even at this earliest epoch. One of the most
prominent is the L1551 region, in the lower left corner of the map. Over the
next 2~Myr, the original centers of activity retained their identity, but new
regions began forming stars. The main characteristic of this evolution
is the increase in star formation along a central group of filaments, lying   
along a Galactic latitude \hbox{$b\,\approx\,-15^\circ$}. This concentration
subsequently grew, so that the youngest stars, with ages under 2~Myr, are    
almost exclusively confined to the filaments. Meanwhile, the total number of 
objects produced per time interval, \ie, the rate of star formation, also rose.
This last trend, of course, is the acceleration we found earlier (Paper II).  

The central filaments represent a concentration not only of stars, but also of
molecular gas. In the top panel of Figure 2, we again display, in addition to
the youngest subset of stars, the $^{12}$CO intensity, now over a more
restricted area. We also include through greyscale shading the distribution of
$^{13}$CO~$(J=1\rightarrow 0)$, taken from Mizuno et al. (1995). For these 
latter observations, the minimum detectable hydrogen column density is
$N_H = 1\times 10^{21}~{\rm cm}^{-2}$. The innermost contours finally display
emission in C$^{18}$O~($J=1\rightarrow 0$) from Onishi et al. (1996). In this
rarest of the three CO isotopes, the minimum column density is
$N_H = 4\times 10^{21}~{\rm cm}^{-2}$. It is apparent that most of the recent 
stellar births are occurring within, or close to, the region bright in
C$^{18}$O. This constitutes, in terms of column density, a narrow ridge      
projecting above surrounding cloud material.

Star formation within the filaments is still active at the present time. This
point is made clear in the bottom panel of Figure 2. Here we reproduce the 
heavy contours of C$^{18}$O from above. We also mark the location of all  
NH$_3$ dense cores that have been found within the borders of the panel  
(Jijina et al. 1999). These 16 local concentrations of cloud gas, detected 
through the (1,1) inversion line, are all associated with C$^{18}$O emission. 
IRAS point sources are found within 10 of the objects.

It is an important characteristic of this system that the sites of individual
protostar collapse are {\it not} scattered more widely throughout the larger 
parent cloud. Let us view the matter from another perspective. Both C$^{18}$O
and NH$_3$ trace especially dense molecular gas. Indeed, the minimum detectable
$N_H$ for the latter is $3\times 10^{21}~{\rm cm}^{-2}$. However, the critical
{\it volume} density for excitation of NH$_3$(1,1) is
$2\times 10^4~{\rm cm}^{-3}$, a factor of 5 above that for
C$^{18}$O~($J=1\rightarrow 0$) (Swade 1989). Thus, while a detection of NH$_3$
essentially guarantees that C$^{18}$O will also be present, the 
converse is false. One can imagine a looser aggregate of NH$_3$ cores, each  
nested within an isolated patch of C$^{18}$O. With the exception of two objects
in the lower right of Figure 2 (called L1489 and L1498), this is not the case.
Most dense cores, along with the new stars they are producing, arise in a  
more contiguous region of enhanced density, \ie, the filaments. 

Two familiar caveats that apply here concern observational selection and
completeness. The surveys for dense cores examined areas already known to have
a high visual extinction (see, e.g., Benson \& Myers 1989), and could well have
missed some objects lying outside the main concentration of cloud gas. In  
addition, the mapping of C$^{18}$O is incomplete. Could there exist other  
star-forming filaments within the area bounded by Figure 1? The contours of   
$^{12}$CO that coincide presently with C$^{18}$O correspond to 
$\int\!T\,dV \gsim 7~K~{\rm km}~{\rm s}^{-1}$. Four other locations satisfy  
this criterion. These include: the L1551 cloud, centered on 
\hbox{$l = 179^\circ, b = -20^\circ$}; part of the Auriga region, near
\hbox{$l = 172^\circ, b = -9^\circ$}; the L1544 cloud at
\hbox{$l = 176^\circ, b = -10^\circ$}; and a smaller region located at
\hbox{$l = 173^\circ, b = -21^\circ$}. Seventeen of our sample T~Tauri stars 
are located in L1551, 7 in the relevant part of Auriga, 6 in L1544, and none 
in the last region. In summary, while the L1551 cloud is the most conspicuous 
in terms of star formation activity, no region, including this one, is
comparable to the main filaments.

Figure 3 compares the ages of stars located inside and outside the filaments,  
where the latter is defined by the C$^{18}$O contours. The two plots reveal
strikingly different star formation histories. Activity was already present
$10^7$~yr ago throughout the cloud. About $3\times 10^6$~yr ago, the rate of  
star formation within the larger, exterior volume reached a modest peak, and
subsequently fell off. Concurrently, the rate within the present C$^{18}$O   
contours attained a much higher level and continues to rise steeply today. 

As we noted earlier, contrasting histories within adjacent regions were also
recently found by Dolan \& Mathieu (2001) in their study of the
$\lambda$~Orionis association. Here, a large evacuated cavity in the ambient
molecular gas surrounds the O star $\lambda$~Ori. Their Figure 11 shows that   
low-mass star formation {\it outside} the cavity, \ie, in the undisturbed 
molecular cloud, is smoothly accelerating. Activity {\it inside} the presently
evacuated central region peaked about $2\times 10^6$~yr ago, fell sharply, and
has by now essentially vanished.

In the case of this OB~association, the recent decline evidently results from
cloud dispersal by a few massive stars. (Recall the discussion of
Upper~Scorpius in Paper II.) In Taurus-Auriga T~association, the decline
throughout the larger volume is less steep, but may still reflect partial and
ongoing dissipation, presumably through the action of low-mass stellar 
outflows. Both regions have vigorous star formation today wherever the parent 
cloud has reached {\it and maintained} a certain density. The recent increase 
in stellar births is so strong in Taurus-Auriga that it gives the impression 
of a {\it global} acceleration when tallying up all stars in the association 
(Paper II). We now see that the rising activity occurs exclusively within the 
central, filamentary region.

\subsection{Nature of the Outlying Stars}

Our account of star formation assumes that {\it all} objects in our list,
including the 80 outside the C$^{18}$O filaments but within the borders of   
Figure 1, are bona fide members of the association. If these outliers were  
instead misclassified foreground stars or interlopers from another region,  
then the Taurus-Auriga history would be quite different. Every object of the 
153 total was classified, in the references cited earlier, as either a  
weak-lined or classical T~Tauri star, on the basis of H$\alpha$ emission,  
proper motion, and radial velocity. In a few cases, lithium abundances were 
available as an independent check (see below). Hence, both the relative youth 
and membership of the outliers appears to be established. 

The last point, that these pre-main-sequence stars were actually born within
the confines of the present-day molecular cloud, is worth emphasizing. In
Figure 4, we show the distribution of stellar radial velocity {\it relative}
to that of the cloud. That is, we plot the difference of $V_\ast$, the stellar
$V_r$-value, and $V_{\rm CO}$, the $^{12}$CO radial velocity at the same
spatial location. The upper panel, which refers to objects inside the
filaments, shows a distribution that is largely symmetric about
\hbox{$V_\ast - V_{\rm CO} = 0$} and extends for at least $\pm 2$~km~s$^{-1}$
on either side. Most of this spread reflects errors in the
measurement of $V_\ast$, which is less precisely determined than the gas
velocity (see, e.g., Hartmann et al. 1986). Thus, the true velocity difference
in each case is almost certainly less. The lower panel, covering objects
outside the filaments, reveals a similar, symmetric pattern. The outliers
appear, therefore, to be physically associated with the molecular cloud.

Are the outliers, on average, older than stars inside the filaments, as Figure
3 indicates? Here a concern is the accuracy of each star's assigned luminosity
and effective temperature. These values determine, via the evolutionary tracks,
the pre-main-sequence contraction age. Paper II gauged the effect of unresolved
binaries, while Hartmann (2001) has recently summarized other sources of error.
For the luminosity, which more directly influences the age, the single biggest
uncertainty arises from patchy interstellar extinction. Fortunately, the
outliers are located in a region of the parent complex with relatively low
column density. Figure 5 illustrates this fact by showing the distribution of
the published $A_V$-values, again both inside and outside the filaments. It is
clear that both the mean visual extinction and the maximal values are   
substantially lower for the outlying stars. Assuming these are located at the
same distance as other association members, their luminosities, and hence ages,
should be relatively secure.

We mentioned earlier that an additional piece of evidence supporting the
pre-main-sequence status of the outliers is enhanced surface lithium
abundance. On the other hand, the mean age of these objects exceeds those
inside the filaments. These two facts might seem to be in contradiction, since 
lithium is consumed during contraction toward the main sequence. The resolution
here is that lithium only disappears early for stars in which the base of the  
convection zone exceeds the ignition temperature of $3\times 10^6$~K (e.g.,   
Siess et al. 2000).

To elaborate on this point, we display in Figure 6 the positions within the  
HR diagram of the 8 outliers for which we have found published lithium  
abundances. Objects inside the white area between the birthline and the ZAMS 
should have the full, interstellar supply of lithium. In the light shaded 
region are objects in which the element has been depleted down to 0.1 times  
the interstellar value (Siess et al. 2000). Finally, the darker shading  
indicates depletion by more than this amount. The dashed curve is the  
isochrone corresponding to a contraction age of $1\times 10^7$~yr. 

Of the 8 objects, the leftmost 3 are on radiative tracks and therefore should
have undepleted lithium, despite ages close to $10^7$~yr. The middle 3 are
slightly younger, and should have modest depletion, while the 2 on the right   
are too young for any lithium consumption. Table 1 gives the observed
abundances in these objects, where each is identified as either a classical (C)
or weak-lined (W) T~Tauri star. Following convention, the figure $N({\rm Li})$
in the fourth column is actually ${\rm log\,(Li/H)}\,+\,12$, where the
argument is the number abundance of lithium relative to hydrogen. For all    
objects, we see that $N({\rm Li})$ is either at its full, interstellar  
value between 2.9 and 3.2 (Zapatero Osorio et al. 2002) or is only modestly  
reduced.

The observational evidence is convincing that the outliers are neither   
interlopers from other star-forming regions nor foreground objects. They
represent a portion of the Taurus-Auriga association which is more widespread
than the generally younger stars within the central filaments. It is unlikely,
moreover, that these objects drifted from the filaments to their present 
locations. In Figure 7, we display proper motion vectors taken from the
literature (Jones \& Herbig 1979; Hartmann et al. 1991; G\'omez et al. 1992;
Frink et al. 1997; Wichmann et al. 2000). Here we have subtracted off the
mean proper motion for the central region, and have omitted all the resulting
vectors whose magnitude is less than the typical error of 5~mas~yr$^{-1}$
(Frink et al. 1997). We see that the velocities of the outliers are {\it not}
directed away from the C$^{18}$O filaments, whose contours we again display.
\footnote{The Taurus-Auriga complex appears in projection within the vast area
covered by the dispersed Cas-Tau association (Blaauw 1991). One might therefore
worry about contamination of our outliers by this older population. Aside from
the lithium and radial velocity tests that distinguish Taurus-Auriga members,
the mean proper motion of the two groups is very different in direction, so that
such contamination should be minimal (de Zeeuw et al. 1999).}

What, then, is the significance of the observed vectors? We emphasize again
that the radial motions show no substantial difference between the velocity of
each star and its local patch of molecular gas. These two velocities must have
been equal just after formation of each object, and they continue to match 
today. The simplest interpretation of this fact is that the stars, which
comprise a tiny fraction of the overall mass, have remained trapped within the
local gravitational potential wells of the gas. The trapping force arises from
parsec-scale regions of the complex, not individual dense cores. Such regions
have their own internal motions; these appear as the vectors in Figure 7. 
Finally, the presence of coherent entities such as L1544 and L1551 also 
attests to the fact that most stars were born in situ. Especially dense clumps
have been producing stars since the earliest days of the complex.  
 
The coexistence of a widespread and spatially concentrated population of young
stars has also been documented in the L1641 region of the Orion~A molecular 
cloud (Strom et al. 1994). Of course, even the peak gas densities in the 
Taurus-Auriga filaments are insufficient to produce massive stars such as those
in Orion. This difference, however, may be one of degree, in a star formation 
pattern that is actually quite similar.

\section{Theoretical Interpretation}

\subsection{Quasi-Static Cloud Evolution}

Our assessment of the ages for Taurus-Auriga members demonstrates that star
formation has been occurring throughout the parent cloud complex for at least
$1\times 10^7$~yr. This figure exceeds by an order of magnitude $t_{\rm coll}$,
the time required for gravitational collapse of the associated molecular gas.
The discrepancy of these two time scales is fundamental in any evolutionary
model, so we should examine it with some care.

Evaluation of $t_{\rm coll}$ proceeds by first estimating the cloud density
through an appropriate tracer. Such a line is  
\hbox{$^{13}{\rm CO}~(J = 1\rightarrow 0)$}, which is optically thin outside
the C$^{18}$O filaments and only marginally thick inside (Mizuno et al. 1995).
Referring again to the upper panel of Figure 2, we selected one representative
location  
\hbox{$(l = 170^\circ, b = -15^\circ)$}  
outside the filaments, but within the region bright in $^{13}$CO. We chose a
second position 
\hbox{$(l = 170^\circ, b = -16^\circ)$}  
wholly within the filaments. Using spectra kindly supplied to us by A. Mizuno, 
we first estimated the column densities $N_H$ in the two locations. Assuming   
the tracer to be optically thin, the $N_H$-values were
$2\times 10^{22}$~cm$^{-2}$ and $9\times 10^{22}$~cm$^{-2}$ for the outer and
inner location, respectively. We then modeled both the $^{13}$CO region and the
ensemble of filaments as uniform cylinders, with radii of 2.2 and 0.35~pc.
Dividing $N_H$ by the corresponding cylinder diameters, we obtained volume 
densities $n_H$. These were 1500~cm$^{-3}$ for the outer point and
4400~cm$^{-3}$ for the inner one. The corresponding values of $t_{\rm coll}$,
taken to be $(3\,\pi/32\,G\,\rho)^{1/2}$, were $9\times 10^5$~yr and  
$5\times 10^5$~yr. For other selected exterior positions, $N_H$ changed by at 
most a factor of 3, so that our $t_{\rm coll}$ is probably secure to within a
factor of 2.

It is interesting to compare $t_{\rm coll}$ with the crossing time of each
region. We may define $t_{\rm cross}$ as $2\,R/\Delta V$, where $R$ is the
cylinder radius and $\Delta V$ the line width (FWHM), as obtained from the
spectra. The crossing times are then $3\times 10^6$~yr and $7\times 10^5$~yr   
for the outer and inner regions. While it would be inappropriate to emphasize
the exact numerical values, it is apparent that $t_{\rm coll}$ and
$t_{\rm cross}$ are quite similar, and that both are larger in the exterior
zone.\footnote{The crossing time would be greater if one used as a length
the major axis of an individual, striated cloud. On the other hand, millimeter
observations show that clumpiness persists even in the direction of greatest
cloud elongation (see, e.g., Figure 1 in Onishi et al. 1998). Our cylindrical
diameters at least crudely represent the spatial scale of the largest clumping.}
A rough match of these two time scales is consistent with the idea that  
the region in question is in force balance. That is, self-gravity is 
effectively opposed by an outward force manifesting itself as an enhanced, 
superthermal line width. This balance is also reflected in the equality of 
cloud {\it masses} derived from optically thin tracers and from the virial 
theorem, an equality that is seen in many systems (e.g., Bertoldi \& McKee 
1992). 

The nature of the outward force is no clearer in Taurus-Auriga than in any     
other molecular cloud complex. The traditional view has been that turbulent  
motion within the strongly magnetized gas supplies the necessary support 
against gravity (e.g., McKee et al. 1993). Within the last few years, this 
view has been challenged, largely as a result of numerical simulations. (For a
review, see V\'azquez-Semadeni et al. 2000.) These studies consider a
computational box filled with gas and a frozen-in magnetic field. When the  
fluid is agitated in a chaotic manner, opposing streams collide and produce  
dense formations within a crossing time. In more detail, it appears that 
transverse (Alfv\'enic) MHD waves of large amplitude become dissipative,
longitudinal disturbances once they travel distances in excess of their own
wavelengths (Hanawa 2001).

While the numerical results are clear enough, their astrophysical implications
are not. The most straightforward interpretation is that all observed molecular
cloud complexes must undergo prompt gravitational collapse. In the case of
Taurus-Auriga, this picture is rendered untenable by the slow evolutionary time
scale, as inferred from the stellar ages. Other dynamical models face a similar
objection (see \S 3.3 below). An alternative view is that the simulations have
failed to capture the true internal motion of the gas, which is actually less
chaotic and more organized. Any transverse wave, for example, apparently must
have a wavelength comparable to the cloud dimension in order to survive. The
motion within the cloud might be better described as a superposition of normal
modes than as a broad spectrum of turbulent eddies.

Whatever its internal dynamics, any cloud destined to form stars must contract
gravitationally. It is tempting to equate the inward motion with that revealed
by asymmetric line profiles in various molecular tracers (Myers et al. 2000).
However, these millimeter observations concern, for the most part, smaller
length scales of order 0.1~pc. Olmi \& Testi (2002) have recently presented   
evidence for cloud contraction over a size of 0.5~pc in Serpens. As these
observations develop, we should bear in mind that such global cloud evolution
must be mediated by energy dissipation. This loss, in turn, arises from shocks,
such as those in the simulations cited previously. A more ordered internal
state implies that the shocks are less pervasive than in a fully turbulent
medium.

\subsection{Star Formation Threshold}

Despite the protracted nature of the cloud contraction, the bulk of star  
formation in Taurus-Auriga took place in the relatively recent past (Paper II).
That is, the appearance of protostars within dense cores is localized in  
{\it time}. The main result of the present study is that this activity is also
localized {\it spatially}, i.e., in the central filaments. These two facts are
undoubtedly related. In the early phase of quasi-static evolution, few places
in the cloud have contracted to the point of forming dense cores and protostars.We witness only scattered activity at isolated locales during this epoch. 
Eventually, contraction produces a substantial, nearly contiguous region with
the requisite density to create cores. At this critical juncture, protostellar
collapse occurs nearly simultaneously throughout the filaments. The spatial and
temporal data, taken together, thus indicate that star formation is a
{\it threshold} phenomenon. More precisely, this threshold refers to the
conditions necessary to form multiple dense cores, which then produce stars on
a relatively brief time scale.

Molecular-line studies of other regions add support to this idea. Kawamura et
al. (1998) conducted a large-scale survey in
\hbox{$^{13}$CO~($J=1\rightarrow 0$)} of the Gemini and Auriga regions. They
identified 139 distinct clouds, with an average size of 3~pc. These regions are
thus comparable to that mapped by the same tracer in Taurus-Auriga. Those 
clouds with embedded infrared sources tend to be more massive and larger.
Conversely, $^{13}$CO clouds without stars have masses only 30 to 50~percent of
the virial value, suggesting that they are partially confined by external
pressure. The most interesting point for our discussion is the existence of a
{\it minimum column density} for the appearance of interior stars. Kawamura
et al. estimate the $N_{{\rm H}_2}$-value as $1.6\times 10^{21}$~cm$^{-2}$,  
corresponding to $N_H\,=\,3\times 10^{21}$~cm$^{-2}$, or an $A_V$ of 1.9~mag. 
A similar result was obtained for the Cepheus-Cassiopeia region by Yonekura et
al. (1997). These authors find a somewhat larger threshold $N_{{\rm H}_2}$ of
$2.5\times 10^{21}$~cm$^{-2}$.

Essentially the same conclusion may be drawn from observations of 
``inefficient"star formation in various complexes. Thus, the Chamaeleon 
star-forming region consists of three main clouds, each with a mass of order 
$10^4\,\Msun$ (Mizuno et al. 2001). Two of the three clouds have embedded 
stars, while Chamaeleon~III has neither infrared sources nor dense cores. This
cloud is actually the most massive, but has the lowest peak surface density, 
as measured in $^{13}$CO (Mizuno et al. 1998). Of the nine clouds in Lupus, 
only two are forming stars. These have the largest $N_H$-values (Hara et al. 
1999).

The example of Chamaeleon~III illustrates that relatively massive molecular
clouds may still be barren of stars. Again, the reason is that the object has
not attained sufficient density. Another example is the Coalsack, a cloud 
of $3500\,\Msun$ at a distance of 180~pc. No infrared sources have been found.
As described by Kato et al. (1999), the $^{13}$CO is almost entirely 
distributed in many small cloudlets less than 0.4~pc in size, These are 
spread widely throughout the parent body, and have an average 
$N_{{\rm H}_2}$ between 1 and $2\times 10^{21}$~cm$^{-2}$, i.e. either at or 
slightly below the putative  threshold. The {\it total} molecular mass  
detected via $^{13}$CO is only 0.2 times that in $^{12}$CO, less than half the
fraction in Taurus-Auriga. Thus, the Coalsack has not produced the analog of 
the central filaments. 

\subsection{Comparison with Dynamical Models}

Our view that clouds evolve relatively slowly contrasts sharply with that
advanced recently by a number of authors. In a widely cited paper, Elmegreen
(2000) has claimed that the duration of star formation in molecular clouds is 
equivalent to one or two crossing times. Recognizing also the equality of 
$t_{\rm cross}$  and $t_{\rm coll}$, Elmegreen argues that those clouds 
destined to form stars undergo prompt, global collapse, and that this collapse
in turn produces the observed stars. 

The observational evidence used to support rapid star formation ranges from the
age difference of Cepheid variables in the LMC to the persistence of
substructure in embedded, Galactic clusters. It is most appropriate to focus on
the arguments based on pre-main-sequence stellar ages. Here, Elmegreen 
considers OB associations, such as the Orion Trapezium region. Unfortunately, 
these are systems in which the massive stars have already driven off the 
parent cloud material, so it is impossible to estimate directly either 
$t_{\rm cross}$ or $t_{\rm coll}$. Even after guessing these times, the 
duration of star formation itself is problematic. As we demonstrated in 
Paper I, most stars within the Trapezium did form in the relatively brief 
interval of $2\times 10^6$~yr. Many other stars, however, preceded them, so 
that activity gradually built up over at least $10^7$~yr, in a manner 
analogous to Taurus-Auriga. Elmegreen correctly points out that continued
equality of the stellar production and crossing time scale would lead to
accelerating star formation in a contracting cloud. For now, our ignorance of
the contraction history in any region prevents verification of such a claim.
In the specific cases of NGC~6611 and NGC~4755, Elmegreen attributes the
observed large age spreads to a series of star formation 
bursts. Our detailed age compilations in Paper II show no evidence for  
intermittent activity in the numerous other systems we examined.

Hartmann et al. (2001) have elaborated on the idea of rapid formation, 
supplying a more detailed physical account. In their picture, a previous 
episode of star formation creates an expanding shell of HI gas. The 
intersection of several shells compresses the gas to the point that it 
collapses, fragmenting simultaneously into stars over a broad front. The 
authors present numerical simulations, similar to those of Ballesteros-Paredes
et al. (1999), to verify that shells are created in large-scale, turbulent 
flows.

The basic idea that new stars may be generated through the external compression
of gas is well supported empirically. HI observations show the presence of    
supershells, hundreds of parsecs in size, that often contain stellar clusters
(Yamaguchi et al. 2001; Elherova \& Palous 2002). On a scale smaller by an
order of magnitude, one sees cloud globules both ablated and compressed by
intense ultraviolet radiation. The densest portion of the globule may include
embedded stars (Sugitani et al. 1995). In both examples, however, the material
is impacted by previously formed {\it massive} stars. There is no evidence that
the molecular clouds in Taurus-Auriga or other sites of low-mass star  
formation were created by this means. On the contrary, the morphology of the   
Taurus clouds, as revealed by high-resolution CO observations, bears little
resemblance to either globules or expanding shells. 

Even if one were to accept that molecular clouds represent swept-up and
colliding shells, the notion that shell collapse produces a stellar association
is ill-founded. A gravitationally unstable slab breaks up into pieces of size
comparable to the original slab thickness (Simon 1965). If one were to identify
each fragment with a star, the breakup would convert essentially all the      
parent molecular gas into stars, in marked contrast to observations. In fact,
there is no basis for such identification. Stars arise from the collapse of
individual cloud cores, not parsec-scale shells or slabs. The buildup of
cores prior to their collapse certainly proceeds faster in a denser cloud 
environment. This density enhancement may be stimulated externally, as in a   
supershell or irradiated globule, or it may occur internally, as in the
central, contracting region of Taurus-Auriga.

\section{Discussion}

Our investigation of Taurus-Auriga has yielded a much more detailed picture 
of star formation in this region than was previously available. We find that   
stellar births occur over a broad area until the cloud's own contraction yields
a region of sufficiently high density to induce much more rapid formation. This
connection between spatial and temporal trends adds further credence to our
pre-main-sequence ages. It also encourages us to apply the same technique to
other sites, in the hope of elucidating fundamental issues concerning stellar
groups. 

The Taurus-Auriga findings already give us further insight concerning the
origin of such associations. Since global contraction has led to the current,
active phase of stellar production, the cloud gas must have been more rarefied
in the past. At that time, the surface density was everywhere below the star
formation threshold. However, with only a modest decrease from the current
$N_H$-value, self-shielding would not have been effective, and the bulk of the
gas would have been HI. An older study by Lucke (1978) may be pertinent in
this regard. Lucke mapped the $B-V$ color excess toward many stars over a large
portion of the Northern sky. He surmised that the Taurus-Auriga molecular      
clouds are actually linked to the Perseus complex through a ``supercloud" of   
HI gas. Whether or not such a coherent structure exists, Lucke's study reminds 
us that the Taurus-Auriga clouds undoubtedly appeared very different prior to
the onset of star formation. The contours in the leftmost panels of our own 
Figure 2, which implicitly assume constancy of the cloud morphology, should  
therefore be viewed only as a crude schematic. 

On the other major question, that of cloud dispersal and the truncation of
stellar births, our study has less to contribute. Most of the 60 stars for   
which we have no estimates are deeply embedded and located in the central 
filaments. Thus, accelerating star formation will undoubtedly continue for  
several Myr. Further progress in understanding the general issue of the cutoff
will come by observing associations slightly more evolved than Taurus-Auriga, 
to witness firsthand expansion of the parent cloud. The discovery of TW~Hydra,
$\eta$~Chamaeleon, and related groups is of considerable interest, but these   
associations are already gas-free. We note finally that expansion will cause   
the H$_2$ gas to become HI, in the reverse process as the prior contraction.   
The sighting of discrete HI clouds with embedded T~Tauri and post-T~Tauri stars
would therefore constitute an important advance. 

\acknowledgments

We are grateful to T. Dame, A. Mizuno, T. Onishi, and J. Swift for supplying 
us with their CO spectra and maps. In addition, R. Cesaroni, F. Massi, and 
L. Testi were of considerable help in the interpetation and analysis of the 
data. F. P. was supported throughout this project by grant ARS~1/R/27/00. 
Funding for S. S. was through NSF Grant AST~99-87266.

\clearpage

\begin{figure}
\plotone{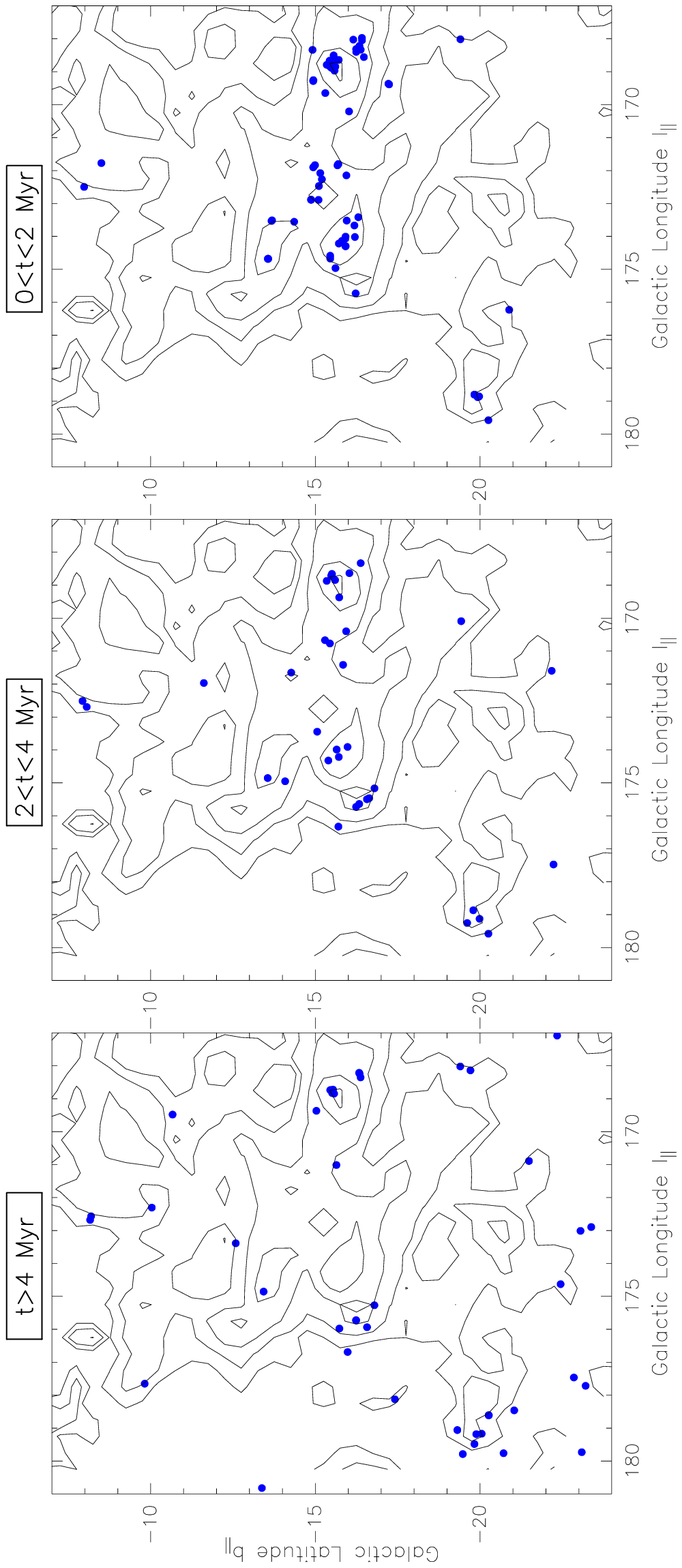}
\caption{
Evolution of the stellar distribution in Taurus-Auriga. The   
dots are observed T~Tauri stars, binned according to pre--main-sequence age, as
shown. Solid contours represent the \hbox{$^{12}$CO~$(J = 1\rightarrow 0)$}
integrated intensity map from Dame et al. (2001). 
\label{fig1}}
\end{figure}

\clearpage 

\begin{figure}
\plotone{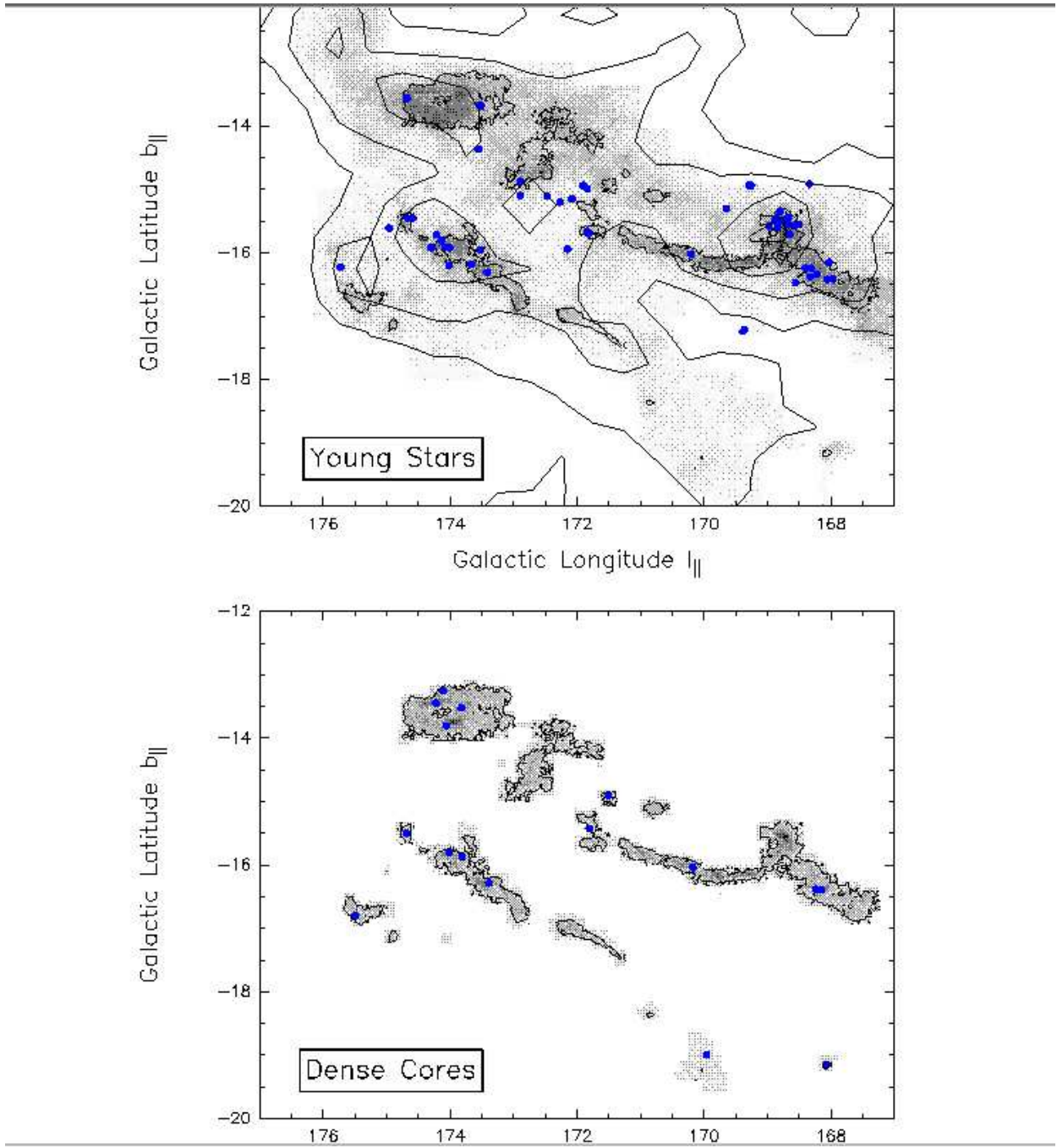}
\caption{
({\it top panel}) Relation of young stars and the central   
filament. The dots are the T~Tauri stars from the rightmost age bin of Figure
1, and the outer contours again show the $^{12}$CO integrated intensity. The
greyscale shading represents $^{13}$CO~\hbox{$(J=1\rightarrow 0$)} flux from
Mizuno et al. (1995). Finally, the innermost contours show
C$^{18}$O~\hbox{($J=1\rightarrow 0$)} from Onishi et al. (1996). ({\it bottom 
panel}) Positions of NH$_3$ dense cores, from Jijina et al. (1999). Also shown
is the C$^{18}$O emission, both as contours and in greyscale.
\label{fig2}}
\end{figure}

\clearpage 

\begin{figure}
\plotone{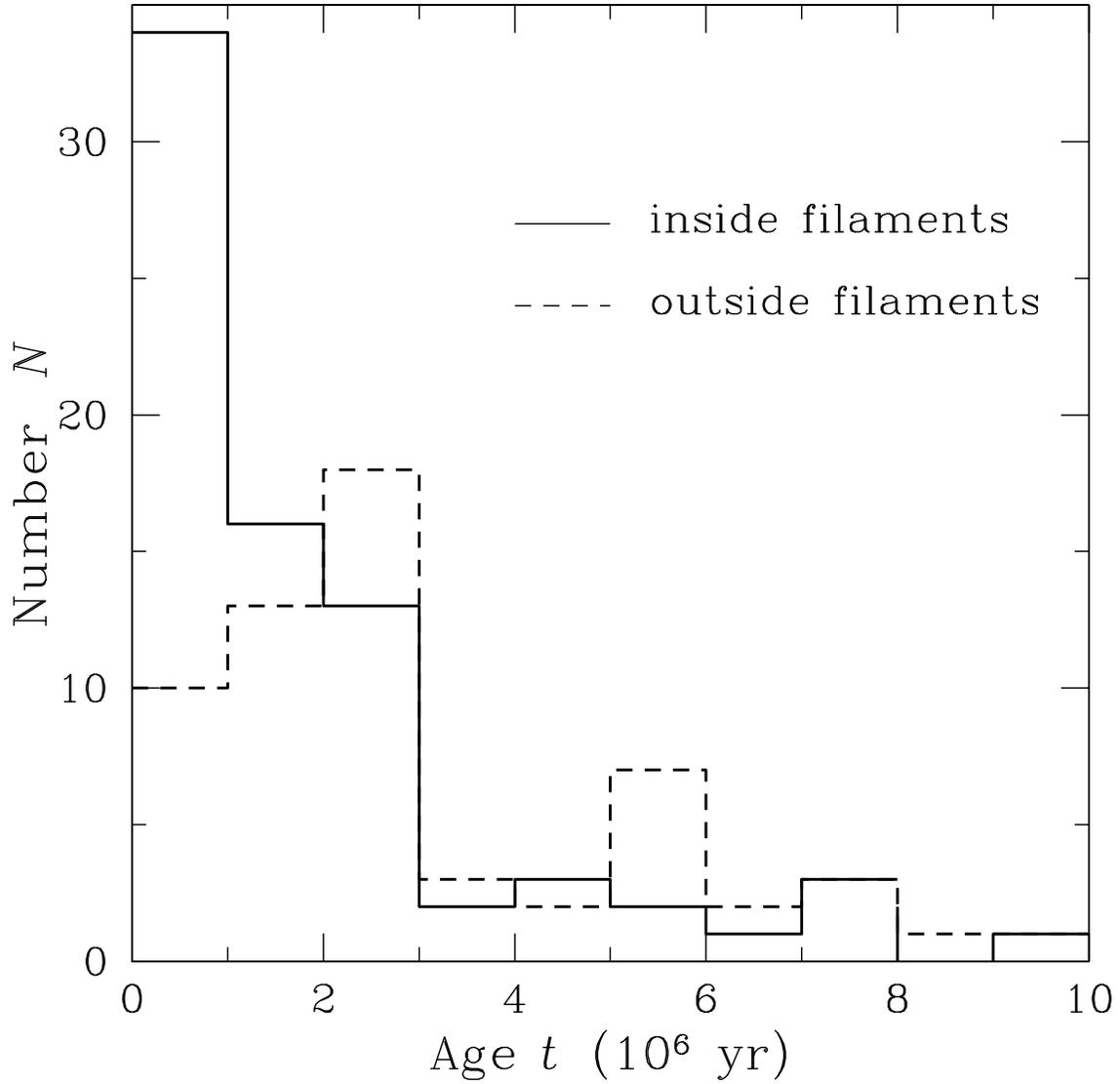}
\caption{
Distribution of pre-main-sequence ages for T~Tauri stars
inside the C$^{18}$O filaments of Figure 2 ({\it solid histogram}). Shown for
comparison ({\it dashed histogram}) are the ages of stars outside the 
filaments, but still within the $^{12}$CO contours of Figure~1.
\label{fig3}}
\end{figure}

\clearpage 

\begin{figure}
\plotone{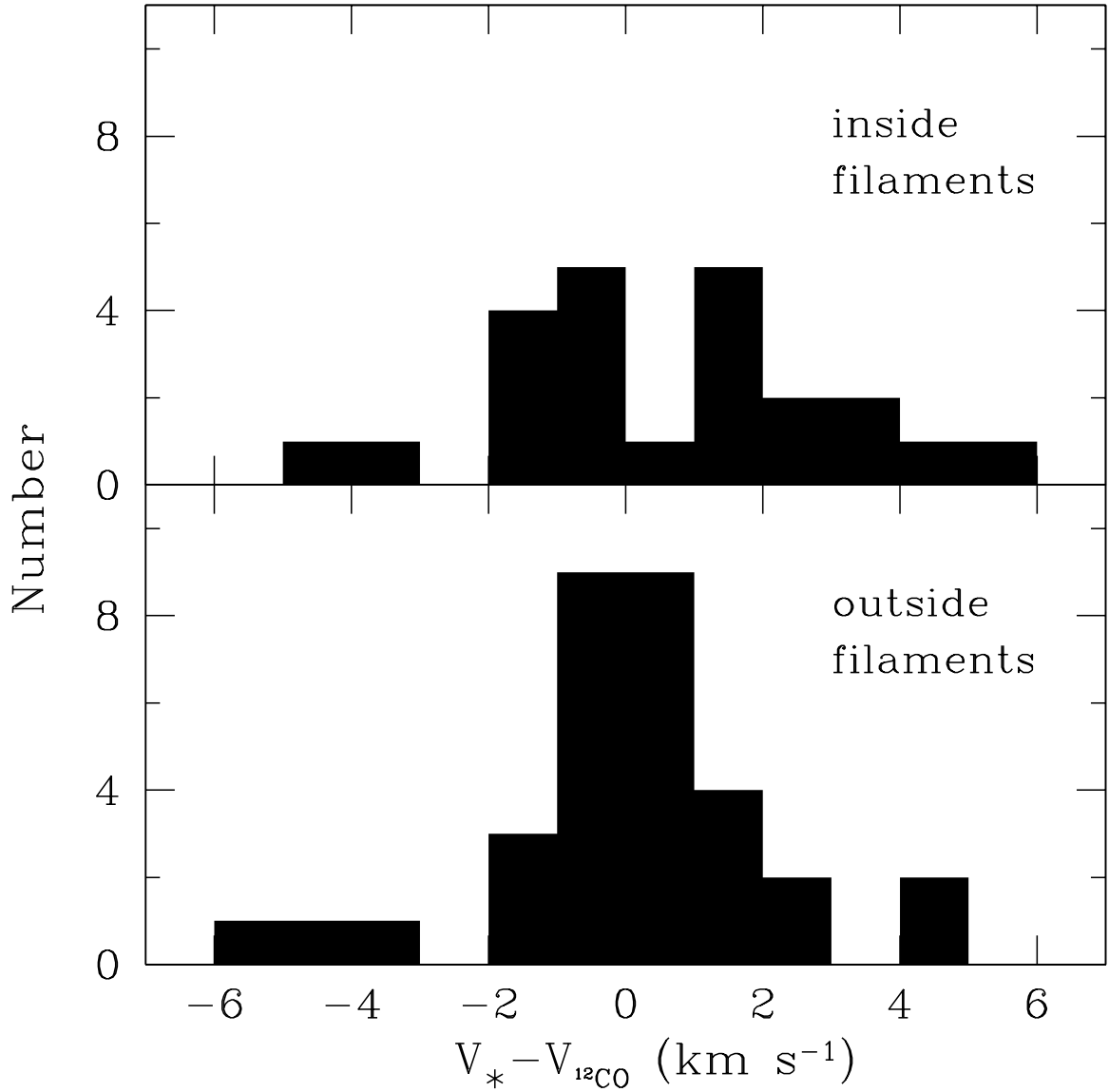}
\caption{
Distribution of the stellar velocity relative to the ambient
cloud, for objects inside and outside the C$^{18}$O filaments. The stellar
velocity is the radial value from the optical spectrum, while the cloud velocityis that obtained in $^{12}$CO~\hbox{($J=1\rightarrow 0$)}.
\label{fig4}}
\end{figure}

\clearpage 

\begin{figure}
\plotone{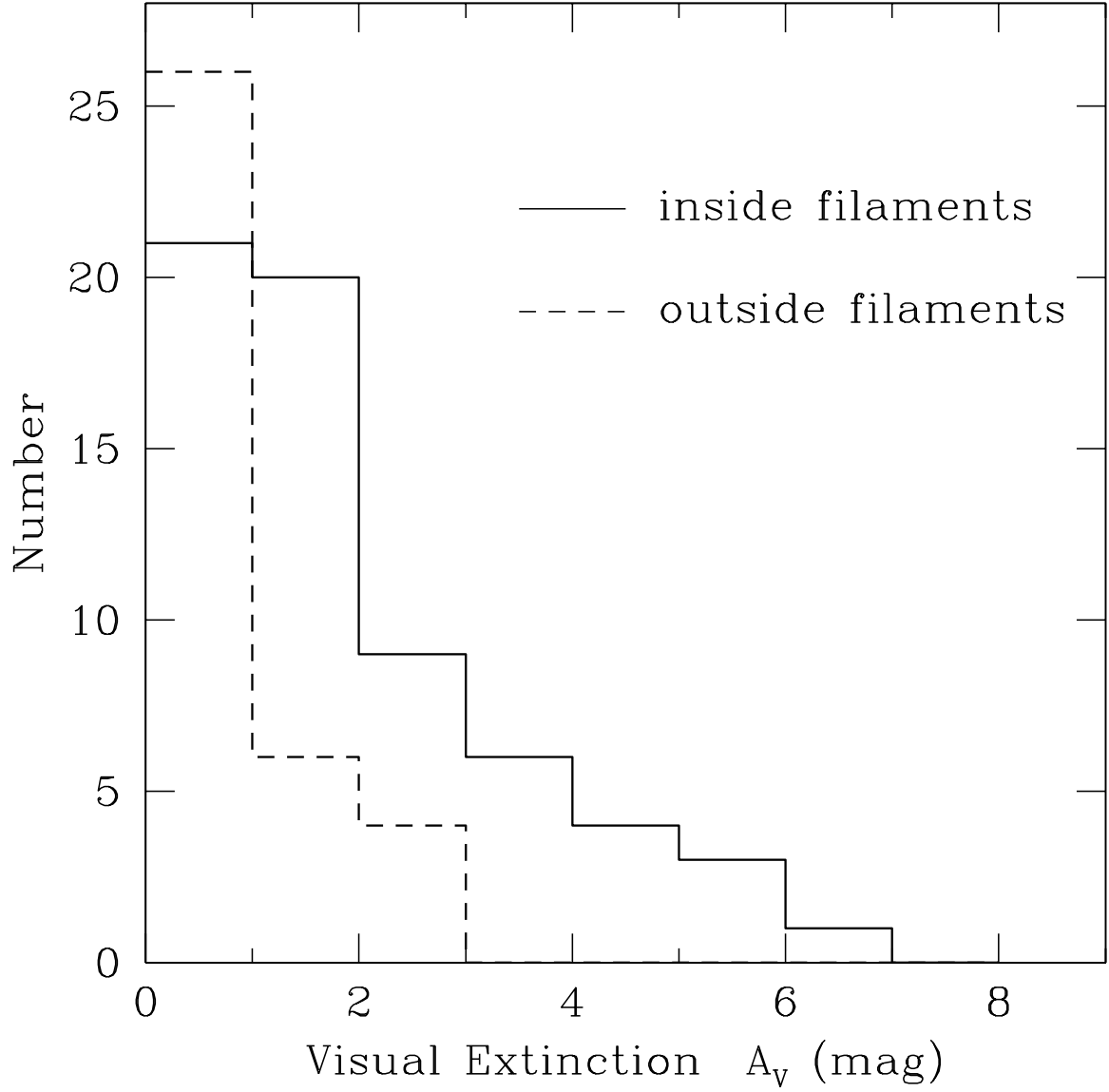}
\caption{
Distribution of the visual extinction $A_V$ for stars both   
inside the C$^{18}$O filaments ({\it solid}) and outside them
({\it dashed}). 
\label{fig5}}
\end{figure}

\clearpage 

\begin{figure}
\plotone{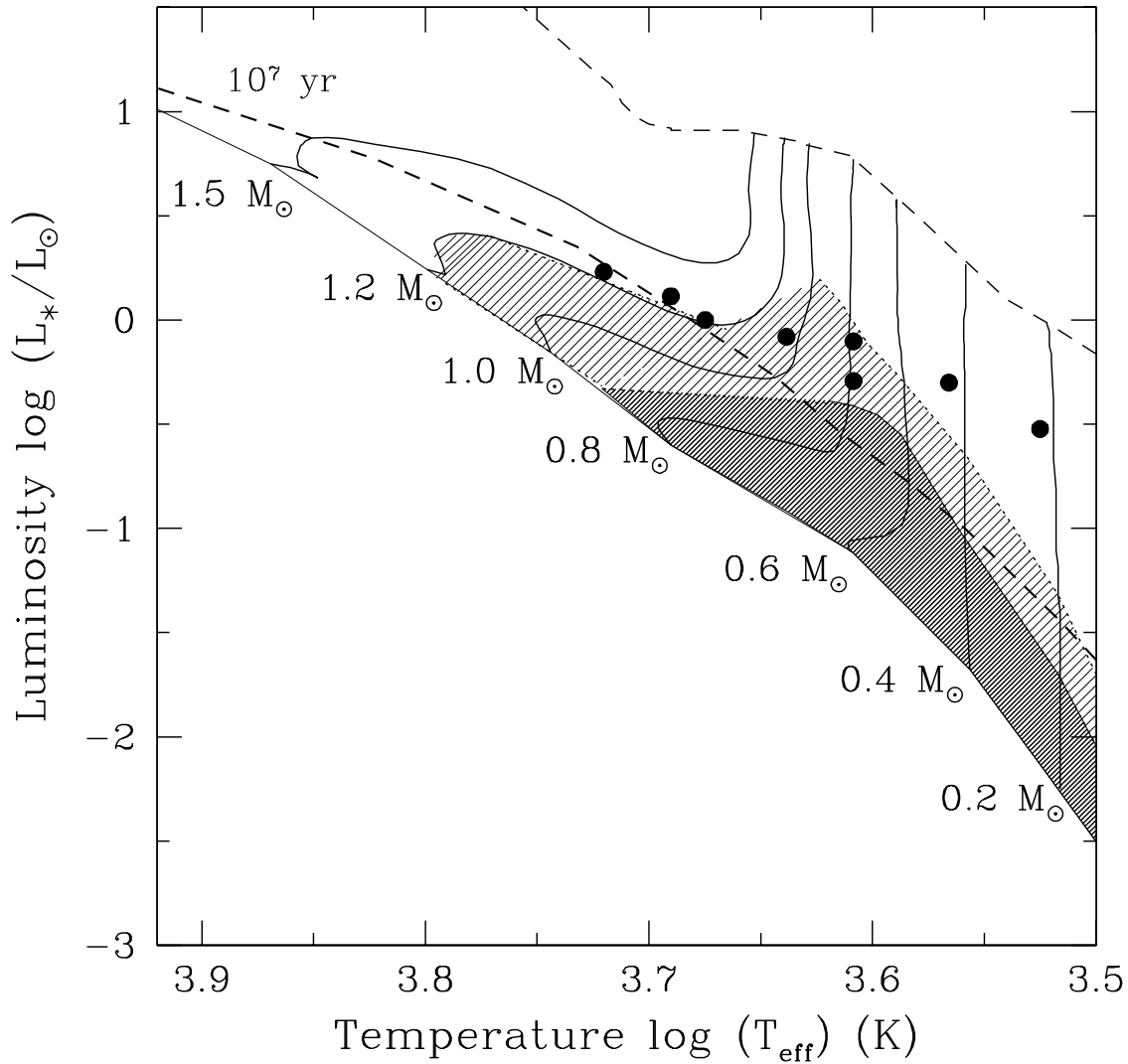}
\caption{
Location in the HR diagram of all outlying stars with measured
lithium abundances. The lighter shaded region represents objects in which
lithium has been depleted as much as 0.1 times the interstellar value,
according to Siess et al (2000). The darker shading is for depletion greater
than this amount. Shown also are pre--main-sequence tracks for the indicated
mass values, in solar units, from Paper I. The dashed curves are the birthline
and the $1\times 10^7$~yr isochrone.
\label{fig6}}
\end{figure}

\clearpage

\begin{figure}
\plotone{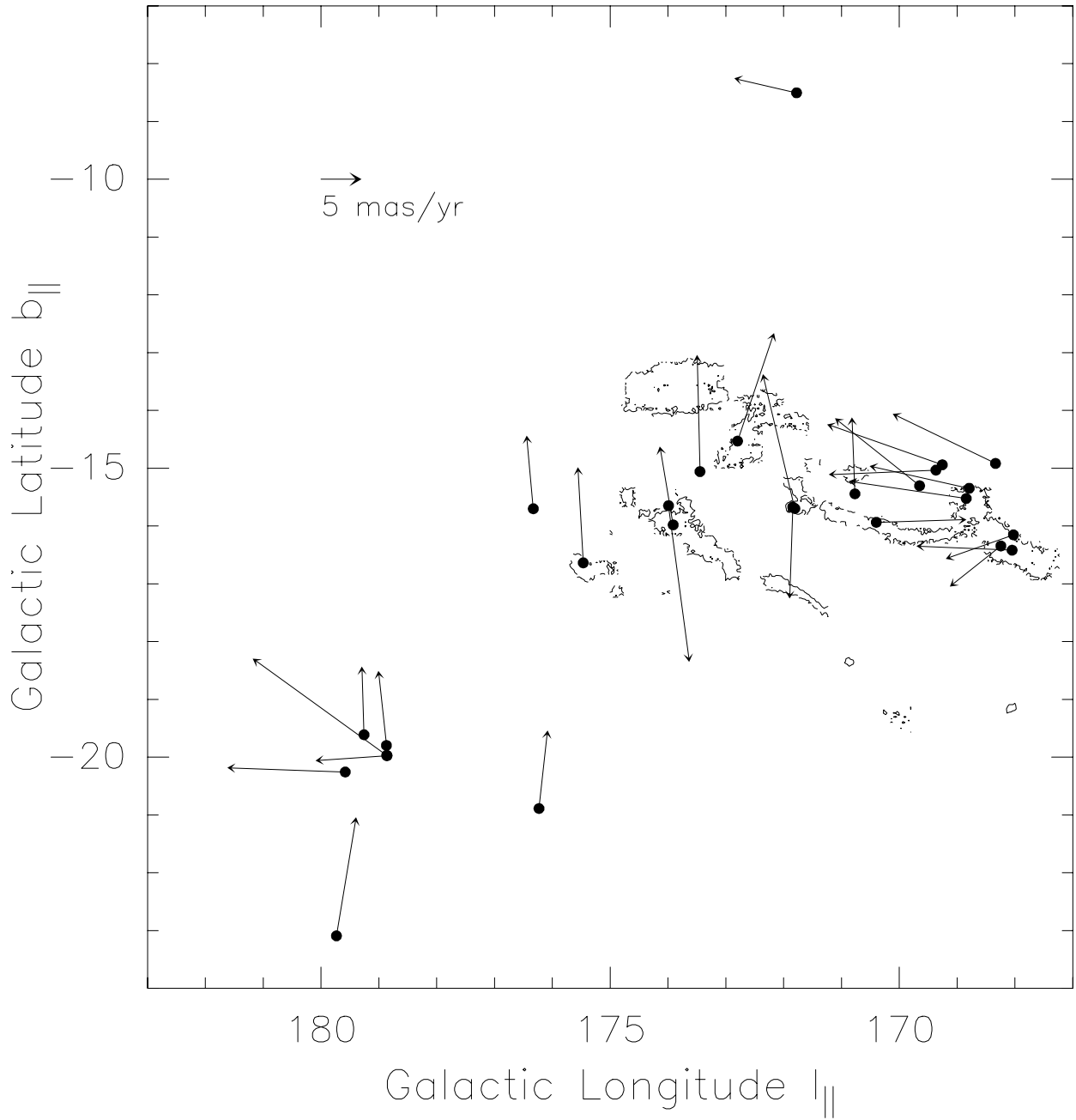}
\caption{
Proper motions of stars in Taurus-Auriga, as obtained from the literature. The
vectors shown are the observed proper motions relative to the mean for the
central filaments (Frink et al. 1997). The C$^{18}$O outer contours are again
superposed. Note that we have omitted all vectors whose magnitude, after 
subtraction of the mean, falls below the typical error of 5~mas~yr$^{-1}$,
which we also display.
\label{fig7}}
\end{figure}

\clearpage

\begin{deluxetable}{lccccc} \footnotesize
\tablecaption{Lithium Abundances in T Tauri Outliers} 
\tablewidth{0pt}\tablehead{
\colhead{Star} & \colhead{Type}   & \colhead{Age}   &
\colhead{$N({\rm Li})$}   & \colhead{Ref.}\\
\colhead{} & \colhead{} & \colhead{(Myr)} & \colhead{} &\colhead{}  
}
\startdata
GM~Aur &W &9.6 &3.1 &1\nl
LkCa~19 &W &14.0 &3.1 &2\nl
LkCa~15 &C &5.2 &3.1 &2\nl
V836~Tau &C &4.9 &3.1 &2\nl
UX~Tau~A &W &8.0 &3.2 &2\nl
LkCa~21 &W &1.5 &3.0 &2\nl
V710~Tau~B &C &0.9 &2.5 &2\nl
V827~Tau &W &2.5 &3.2 &2\nl
\enddata
\tablerefs{
(1) Magazz\'u et al. (1992); 
(2) Mart\'{\i}n et al. (1994)}
%\tablenotetext{}{1:~Magazz\'u et al. (1992)}
%\tablenotetext{}{2:~Mart\'{\i}n et al. (1994)}
\end{deluxetable}

%% The following command ends your manuscript. LaTeX will ignore any text
%% that appears after it.

\end{document}